\documentclass[12pt]{article}
\usepackage{amsmath,amssymb,amsfonts}
\usepackage[dvips]{graphicx}
\usepackage{epsfig}

\makeatletter \@addtoreset{equation}{section} \makeatother

\begin{document}

\begin{titlepage}

\begin{center}
\hfill KIAS-P06030\\
\hfill hep-th/0607085

\vspace{2cm}

{\Large\bf $\frac{1}{16}$-BPS Black Holes and Giant Gravitons

\vskip 0.2cm

in the AdS$_5\times$ S$^{5}$ Space}

\vspace{1.5cm}

{\large Seok Kim$^{a}$ and Ki-Myeong Lee$^{b}$}

\vspace{.7cm}

{\it School of Physics, Korea Institute for Advanced Study,
Seoul 130-722, Korea.}\\

\vskip 0.5cm

{e-mail:\ \ seok@kias.re.kr $^a$,\ \ klee@kias.re.kr $^b$}

\end{center}

\vspace{1.5cm}

\begin{abstract}

We explore $\frac{1}{16}$-BPS objects of type IIB string theory in
$AdS_5\times S^5$. First, we consider supersymmetric $AdS_5$ black
holes, which should be  $\frac{1}{16}$-BPS and have a characteristic
that not all physical charges are independent. We point out that the
Bekenstein-Hawking entropy of these black holes admits a remarkably
simple expression in terms of (dependent) physical charges, which
suggests its microscopic origin via certain Cardy or Hardy-Ramanujan
formula. We also note that  there is an upper bound for the angular
momenta given by the electric charges. Second, we construct a class
of $\frac{1}{16}$-BPS giant graviton solutions  in $AdS_5\times S^5$
and explore their properties. The solutions are given by the
intersections of $AdS_5\times S^5$ and complex 3 dimensional
holomorphic hyperspaces in $\mathbb{C}^{1+5}$, the latter being the
zero loci of three holomorphic functions which are homogeneous with
suitable weights on coordinates. We investigate examples of giant
gravitons, including their degenerations to tensionless strings.

\end{abstract}

\end{titlepage}

\tableofcontents

\section{Introduction}

The type IIB superstring theory on $AdS_5\times S^5$ contains a
large class of BPS states in its spectrum. In the $\frac{1}{2}$-BPS
sector, the BPS states carry nonzero $U(1)\subset SO(6)$ momenta or
the R-charges and have been extensively studied, both in the string
theory and the ${\mathcal{N}}=4$ Yang-Mills theory. With large
R-charges, gravitons expand into spherical $D3$ branes in $S^5$ or
$AdS_5$, called the giant gravitons \cite{mst,gmt,hhi,bbns,cjr}.
More recently some pioneering works in the supergravity and the
super Yang-Mills theory have been done in \cite{llm,ber-1}, as well
as many others. Just to mention a few of them,
\cite{surya,sh-to,mandal,gmmpr,dms}.

Down to $\frac{1}{8}$-BPS sector, BPS states carrying three
R-charges in $U(1)^3\subset SO(6)$ have also been studied.
Especially, in the probe limit, the classical configurations are
given by $D3$ branes expanding in $S^5$ with their shapes given by
holomorphic surfaces in ${\mathbb{C}}^3$ \cite{mikh}. See also
\cite{kl,ber-2,bglm,ma-sur,ast} for related works in the
$\frac{1}{8}$-BPS sector. The same sector has also been investigated
from the supergravity viewpoint. Supergravity solutions carrying
$U(1)^3$ charges in the internal $S^5$ are called superstars
\cite{my-ta,bcs}, which develop naked singularities. These solutions
are interpreted in \cite{my-ta} as distributions of giant gravitons.

It would be interesting to consider the even less supersymmetric
sector, the $\frac{1}{16}$-BPS one preserving 2 real
supersymmetries. One motivation for this comes from the study of
supersymmetric $AdS_5$ black holes, which have been obtained rather
recently \cite{gu-re-1,gu-re-2,cclp1,cclp2,klr} from 5 dimensional
gauged supergravity theories. These black holes carry three
$U(1)^3\subset SO(6)$ momenta in $S^5$ as well as two $U(1)^2\subset
SO(4)$ in $AdS_5$. Note that, in order to have black holes with
regular horizons, the angular momenta in $AdS_5$ should be nonzero:
otherwise the solutions would develop naked singularities
\cite{my-ta,bcs}. The presence of angular momenta in $AdS_5$ forces
them to preserve no larger than $\frac{1}{16}$ supersymmetry. To
understand these objects, it will also be interesting to consider
the microscopic side of this sector and try to compare it with the
supersymmetric black holes. The purpose of this paper is to explore
the macroscopic and microscopic aspects of the $\frac{1}{16}$-BPS
objects, hoping that our results would provide important clues for
future works toward relating both sides: for instance, counting the
black hole microstates. For the previous attempts to understand
microscopic aspects of these black holes, see \cite{kmmr,brs}. (See
\cite{silva} also.)

From the macroscopic side, we consider the Bekenstein-Hawking
entropy of supersymmetric $AdS_5$ black holes. As we already
mentioned, they carry two angular momenta in $AdS_5$, call it $J_1$
and $J_2$, as well as three $U(1)^3\subset SO(6)$ charges $Q_I$
($I=1,2,3$). Another odd feature of the supersymmetric $AdS$ black
holes (known to date) is that these physical charges are not all
independent in order to have regular black hole
solutions.\footnote{See \cite{gu-re-1} for some arguments on this
issue for black holes in minimal gauged supergravity, and
\cite{cglp,kp} for examples of supersymmetric $AdS_7$ and $AdS_4$
black holes with the same property. We are not sure whether this
phenomenon is an essential or a technical one. However, see
\cite{brs}.} Therefore, there should be an ambiguity if one tries to
write down the expression for the entropy in terms of (dependent)
physical charges, which one usually does in order to compare it to
the microscopic one. Indeed, in the literatures, implicit
prescriptions are made on this ambiguity \cite{kmmr,brs} in special
regimes of charges. We observe that, in terms of dependent physical
charges, the entropy admits the following simple expression,
\begin{equation}
  S=2\pi\sqrt{Q_1Q_2+Q_2Q_3+Q_3Q_1-N^2 J}\ ,
\end{equation}
where $J=\frac{J_1+J_2}{2}$ is the self-dual part of the angular
momentum. (See section 2 for the details and generalization.)
Compared to the expressions advocated in previous works, this
expression is exact in all regime of charges. The factor $N^2$
originates from the central charge of the $\mathcal{N}=4$ Yang-Mills
theory which is holographically dual to the string theory in
$AdS_5\times S^5$: $4c=N^2$. We think the expression is remarkably
simple that there should be a nice microscopic explanation, for
instance, via a Cardy or Hardy-Ramanujan formula of a certain
microscopic model.

The above observation would be a motivation for investigating the
$\frac{1}{16}$-BPS sector microscopically. In this diretion, we
directly generalize the work of \cite{mikh} to construct
$\frac{1}{16}$-BPS $D3$ giant graviton solutions in the probe limit
and study their properties. Turning off the worldvolume gauge
fields, we find that $\frac{1}{16}$-BPS brane embeddings and time
evolutions are given by the intersectionss of holomorphic
3-manifolds in $\mathbb{C}^{1+2}\times \mathbb{C}^3$ with
$AdS_5\times S^5$. With a suitable choice of complex structure of
$\mathbb{C}^{1+2}\times\mathbb{C}^3$ with coordinates $Y^A$
($A=1,\cdots,6$), the 3-manifold is given by the zero locus of three
holomorphic functions $F_k(Z^A)\!=\!0$ ($k\!=\!1,2,3$), where the
functions should be homogeneous in the following sense,
\begin{equation}
  F_k(\lambda  Y^0, \lambda Y^1,\lambda Y^2, \lambda^{-1} Y^3,\lambda^{-1} Y^4,\lambda^{-1} Y^5)
  =\lambda^{d_k}F(Y^A)
  \end{equation}
with suitable degrees $d_k$. It will be interesting to construct
explicit examples which are relevant to the black hole physics. We
do not have much to say about it so far, except for a few comments
in section 5.

As an aside, we use these giant graviton solutions to illustrate how
extended objects (like strings) in $AdS_5\times S^5$ would expand to
$D3$ branes, just as the point-like gravitons do. A related issue
has been addressed in \cite{mmt1}: in that paper, it was conjectured
that nearly-BPS ultra-relativistic strings with large angular
momenta $Q_I\gtrsim \sqrt{N}$ in $S^5$ would expand into giant
gravitons. We show that a BPS cousin of this phenomenon, somewhat
similar to the above, would appear in $\frac{1}{4}$-, $\frac{1}{8}$-
and $\frac{1}{16}$-BPS sectors. We first show that there are limits
where $D3$ giant gravitons shrink to string-like configurations. In
some simple cases (to be explained below), the shinking D3 brane is
tubular with topology $S^1\times S^2$, where $S^2$ is small while
$S^1$ remains macroscopic. The profiles of $S^1$ agree with those of
the tensionless string solutions of \cite{mmt2} and generalizations
thereof.

The organization of this paper is as follows. In section 2 we report
our observation on the entropy of supersymmetric $AdS_5$ black
holes. In section 3 we derive a class of $\frac{1}{16}$-BPS giant
graviton solutions in $AdS_5\times S^5$. In section 4 we present
some examples. Especially, we point out the existence of tube-like
solutions with tensionless string limits, analogous to the
point-like graviton limit of spherical giant gravitons. We conclude
in section 5 with several remarks. In the appendix, we generalize
the tensionless string solutions of \cite{mmt2} to the
$\frac{1}{16}$-BPS sector.

As we were preparing this paper, we recognized that section 3
overlaps with some results presented by Shiraz Minwalla in Strings
2006, Beijing. (See also \cite{bglm}.)

\section{Entropy of supersymmetric AdS$_5$ black holes}

Recently a large class of supersymmetric black hole solutions are
discovered in 5 dimensional gauged supergravity
\cite{gu-re-1,gu-re-2,cclp1,cclp2,klr}, preserving 2 real
supersymmetries. Although our main interest is the black holes in
$AdS_5\times S^5$, we start from more general supergravity theories,
following \cite{gu-re-2}. We consider 5 dimensional
${\mathcal{N}}\!=\!1$ gauged supergravity coupled to $n$ abelian
vector multiplets. The bosonic fields are metric $g_{\mu\nu}$, one
graviphoton plus $n\!-\!1$ vector fields collected together as $A^I$
($I=1,\cdots,n$), and $n\!-\!1$ real scalars $\phi^a$
($a=1,\cdots,n\!-\!1$). One introduces $n$ scalars $X^I(\phi^a)$
constrained as
\begin{equation}\label{real special 1}
  \frac{1}{6}C_{IJK}X^IX^JX^K=1\ ,
\end{equation}
where $C_{IJK}$ are constants with symmetric $IJK$ indices. We also
define
\begin{equation}\label{real special 2}
  X_I\equiv\frac{1}{6}C_{IJK}X^JX^K\ \ \
  {\rm which\ satisfy}\ \ \ X_IX^I=1\ .
\end{equation}
The bosonic part of this supergravity action is \cite{gu-sa}
\begin{equation}
  S=\frac{1}{16\pi G}\int\left(
  R^5-2\chi^2{\mathcal{V}}-Q_{IJ}F^I\wedge\ast F^J-Q_{IJ}dX^I\wedge
  \ast dX^J-\frac{1}{6}C_{IJK} A^I\wedge F^J\wedge F^K
  \right)\ ,
\end{equation}
with the coupling matrix $Q_{IJ}$
\begin{equation}
  Q_{IJ}=\frac{9}{2}X_IX_J-\frac{1}{2}C_{IJK} X^K
\end{equation}
and the scalar potential ${\mathcal{V}}$
\begin{equation}\label{potential}
  {\mathcal{V}}=\frac{9}{2}V_IV_J\left(Q^{IJ}-2X^IX^J\right)\ ,
\end{equation}
where $Q^{IJ}$ is the inverse matrix of $Q_{IJ}$. $V_I$ is a
constant vector, which is related to the vacuum value $\bar{X}_I$ of
the scalars $X_I$,
\begin{equation}
  \bar{X}_I=\xi^{-1}V_I
\end{equation}
with a constant $\xi$. The latter constant is fixed by the vacuum
value of the potential:
\begin{equation}
  2\chi^2{\mathcal{V}}(\bar{X}^I)\equiv
  -12\chi^2\xi^2\equiv-\frac{12}{\ell^2}\ .
\end{equation}
The parameter $\ell$ is the radius of $AdS_5$ (supposing that
${\mathcal{V}}(\bar{X}^I)$ is not zero).

For the $U(1)^3$ $\subset SO(6)$ truncation of ${\mathcal{N}}\!=\!4$
gauged supergravity, which can be embedded into type IIB string
theory in $AdS_5\times S^5$ \cite{cvetic}, one has three vector
fields and
\begin{equation}\label{S5 constant}
  C_{123}=1\ \ \ ({\rm other}\ \ C_{IJK}{\rm s\ are\ zero})\ .
\end{equation}
The three constrained scalars $X^I$, measuring the squashing of
$S^5$, take vacuum values
\begin{equation}\label{S5 vev}
  \bar{X}^I=1\ \ (I=1,2,3)\ \ \rightarrow\ \ \
  \bar{X}_I=\frac{1}{3}\ .
\end{equation}
When the scalars take these values, the internal $S^5$ becomes a
round sphere. At any stage of our following analysis, inserting the
above values will give us the results on black holes in $AdS_5\times
S^5$.

In the supergravity considered in \cite{gu-re-2,klr}, the scalars
$\phi^a$ live on a \textit{symmetric space}, where the latter is
specified by the following condition
\begin{equation}\label{symmetric}
  C^{IJK}C_{J(LM}C_{PQ)K}=\frac{4}{3}\delta^{I}_{(L}C_{MPQ)}
\end{equation}
with some $C^{IJK}$. For the $S^5$ case, this requirement is met by
setting $C^{IJK}=C_{IJK}$ with (\ref{S5 constant}). We briefly
comment on other possible symmetric spaces in section 5.

We consider the black holes in the above supergravity theory. For
simplicity, we first consider the black holes with self-dual angular
momentum in $AdS_5$ \cite{gu-re-2}, and comment on more general ones
in \cite{klr} afterwards. Referring to \cite{gu-re-2} for the
details, here we summarize the physical quantities. The $U(1)^n$
charges $Q_I$ are as follows (the normalization of the charges are
commented below):
\begin{equation}\label{electric charge}
  Q_I=\frac{\pi\ell}{G}\left(
  \frac{3}{4}q_I-\frac{3\alpha_2}{8\ell^2}\bar{X}_I+
  \frac{9}{8\ell^2}C_{IJK}\bar{X}^J C^{KLM}q_L q_M\right)\ ,
\end{equation}
where $q_I$ ($I=1,\cdots, n$) are the $n$ independent parameters of
this solution. Defining
\begin{equation}\label{para relation}
  \alpha_1\equiv\frac{27}{2}C^{IJK}\bar{X}_I\bar{X}_J q_K\ ,\ \
  \alpha_2\equiv\frac{27}{2}C^{IJK}\bar{X}_I q_J q_K\ ,\ \
  \alpha_3\equiv\frac{9}{2}C^{IJK}q_I q_J q_K\ ,
\end{equation}
the self-dual angular momentum is
\begin{equation}
  J\equiv\frac{J_1+J_2}{2}=
  \frac{\pi}{8G\ell}\left(\alpha_2+\frac{2\alpha_3}{\ell^2}\right)
\end{equation}
where $J_1$, $J_2$ are Cartans of the $SO(4)$ rotation symmetry of
$AdS_5$. The mass is given as
\begin{equation}\label{AD mass}
 M=\frac{\pi}{4G}\left(\alpha_1+\frac{3\alpha_2}{2\ell^2}+
 \frac{2\alpha_3}{\ell^4}\right)=
 \frac{1}{\ell}\left(\bar{X}^I Q_I + J_1+J_2\ \right)\ .
\end{equation}
$\bar{X}^I$ is the asymptotic value of the scalar $X^I$ in this
solution, the minima of the potential (\ref{potential}). The
Bekenstein-Hawking entropy of the black hole is
\begin{equation}\label{3-para entropy}
  S_{BH}=\frac{A_{S^3}}{4G}=\frac{\pi^2}{2G}
  \sqrt{\alpha_3\left(1+\frac{\alpha_1}{\ell^2}\right)-
  \frac{\alpha_2^2}{4\ell^2}}\ ,
\end{equation}
given by the area of the horizon (squashed 3-sphere).

There are $n+1$ independent physical charges carried by this black
hole: $n$ electric charges $Q_I$'s and the self-dual angular
momentum $J=\frac{J_1+J_2}{2}$. However, there are only $n$
independent parameters $q_I$ of the solution. Thus, there is one
relation between these charges. If one tries to express the
macroscopic entropy in terms of physical charges $Q_I$ and $J$,
there should be an ambiguity in its expression \cite{kmmr,brs} as we
mentioned in the introduction. We take advantage of this ambiguity
and try to write (\ref{3-para entropy}) in terms of the physical
charges in a simple way.

After some trials and errors, we find that the combination
$C^{IJK}\bar{X}_IQ_J Q_K$ is interesting. Using (\ref{real special
1}), (\ref{real special 2}), (\ref{para relation}) and the symmetric
space condition (\ref{symmetric}), one obtains the following result
after some algebra:
\begin{equation}\label{quadrature}
  \frac{3}{2}C^{IJK}\bar{X}_IQ_J Q_K=\frac{1}{16}
  \left(\frac{\pi\ell}{G}\right)^2 \left[\alpha_2+\frac{3\alpha_3}{\ell^2}+
  \frac{\alpha_3\alpha_1}{\ell^4}-\frac{\alpha_2^{\ 2}}{4\ell^4}\right]=
  \left(\frac{S_{BH}}{2\pi}\right)^2+\left(\frac{\pi\ell^3}{2G}\right) J\ .
\end{equation}
Defining the constants
\begin{equation}\label{central}
  c\equiv\frac{\pi\ell^3}{8G}\ \ ,\ \ \
  D^{IJ}\equiv\frac{1}{2}C^{IJK}(3\bar{X}_K)\ ,
\end{equation}
one obtains
\begin{equation}\label{entropy 2}
  S_{BH}=2\pi\sqrt{D^{IJ}Q_J Q_K- 4cJ}\ .
\end{equation}
The constant $c$ defined as (\ref{central}) is actually the central
charge of the holographically dual 4d superconformal field theory
computed from the gravity data, normalized as $c=\frac{N^2}{4}$ for
the ${\mathcal{N}}=4$ $SU(N)$ Yang-Mills theory.

The convenient normalization convention of the electric charges
$Q_I$ depends on the way we embed this gauged supergravity into
higher dimensional string or M-theories. Here we normalized them so
that
\begin{equation}\label{mass contribution}
  \frac{1}{\ell}\bar{X}^IQ_I
\end{equation}
becomes the electric charge contribution to the BPS mass (\ref{AD
mass}). This normalization is natural in the $AdS_5\times S^5$ case,
since $Q_I$ are internal Kaluza-Klein momenta associated with
$U(1)^3\subset SO(6)$ isometry, normalized as integers in our
convention. In general, one may consider two kinds of electric
charges. First, for those associated with internal isometries, the
$\ell$ factor in front of (\ref{electric charge}) should be replaced
by `internal radii' in order to make them integral. In this case,
the microscopic objects carrying these charges would be the (giant)
gravitons. Second, if the internal manifold contains topological
cycles, one can also have charged objects from wrapped branes. In
this case, to define $Q_I$'s as integral wrapping numbers, one needs
to re-scale (\ref{electric charge}) such that $1/\ell$ in (\ref{mass
contribution}) is replaced by the volume times tension factor of the
wrapped branes.

For the $AdS_5\times S^5$ case, or the $U(1)^3$ supergravity, the
charges $Q_I$ are integral as explained above. Inserting the values
of the constants, (\ref{S5 constant}) and (\ref{S5 vev}), the
Bekenstein-Hawking entropy (\ref{entropy 2}) takes the form
\begin{equation}\label{entropy 3}
  S=2\pi\sqrt{Q_1Q_2+Q_2Q_3+Q_3Q_1-N^2 J}\ .
\end{equation}
We again emphasize that the above expression is not unique from the
black hole solutions we have, due to a relation of physical charges
$Q_I$ and $J$. We just discovered a simple way of writing it. For
instance, \cite{kmmr} considered the `small' black holes in the
regime $Q_I\ll N^2$. The relation between the physical charges
$Q_I$, $J$ is
\begin{equation}\label{charge relation}
  N^2 J\approx Q_1Q_2+Q_2Q_3+Q_3Q_1
  -\frac{2}{N^2}Q_1Q_2Q_3+O\left(\frac{Q_I}{N^2}\right)^4
\end{equation}
where all $Q_I$'s are assumed to be of same order. (\ref{entropy 3})
can be rewritten as
\begin{equation}\label{entropy 4}
  S\approx 2\sqrt{2}\pi\frac{\sqrt{Q_1Q_2Q_3}}{N}\ +\
  ({\rm higher\ order\ terms })\ .
\end{equation}
This form (\ref{entropy 4}) was advocated by the authors of
\cite{kmmr} to discuss the microscopic aspects. No matter what
prescription one gives to this ambiguity, the \textit{value} of the
entropy is the same, of course, upon imposing the relation like
(\ref{charge relation}). However, we think (\ref{entropy 3}) would
be significant since this `simple' expression is exact in any regime
of charges. It is likely that there would be a simple explanation of
this degeneracy from the microscopic side, probably via a Cardy of
Hardy-Ramanujan formula in a certain microscopic model. If this
conjecture is true, it will be challenging to explain the
$4c\!=\!N^2$ factors in (\ref{entropy 3}).

From the expression (\ref{entropy 3}), one finds that one of the
three charges may be turned off, say $Q_3\!=\!0$, while having
regular black holes. However, it is impossible to have black holes
with single electric charge, say $Q_2\!=\!Q_3\!=\!0$, since the
quantity inside the square-root cannot be positive any more. This
fact was also mentioned in \cite{gu-re-2}. Saying it differently,
the self-dual angular momentum $J$ of the black hole is bounded from
above by the electric charges $Q_I$. From (\ref{entropy 3}), or
(\ref{entropy 2}), the bound is
\begin{equation}\label{angular bound}
  J_1+J_2\leq\frac{1}{2c}\ D^{IJ}Q_IQ_J\ \ \ \
  {\rm or}\ \ \ \ J_1+J_2\leq \frac{2}{N^2}
  (Q_1Q_2+Q_2Q_3+Q_3Q_1)\ .
\end{equation}
Existence of such an upper bound itself is familiar from the
spinning black holes in the Minkowski space, perhaps except for the
central charge factor $N^2$.

Finally, we turn to the black holes with general angular momenta
$J_1\!\neq\!J_2$. The black holes in \cite{klr} carry $n+2$ physical
charges $Q_I$, $J_1$ and $J_2$ and $n\!+\!1$ independent parameters
in the solutions (generalizing $q_I$'s above). There again is one
relation between physical charges. Compared to the $J_1\!=\!J_2$
solution explained in this section, there is one more parameter and
physical charge, respectively. The way this additional parameter
appears is rather involved, which is the reason why we do not
present it here. Quite remarkably, we find that the simple
expressions (\ref{entropy 2}) and (\ref{entropy 3}) continue to
hold.

\section{$\frac{1}{16}$-BPS giant gravitons}

In this section we find $\frac{1}{16}$-BPS giant graviton solutions
in terms of three holomorphic functions and discuss their
properties.

The solution is given by the six complex coordinates $Y_A$
($A=0,\cdots,5$) of $AdS_5\times S^5 \subset
{\mathbb{R}}^{2+4}\times {\mathbb{R}}^6$, which satisfy the
constraints
\begin{equation}\label{constraint}
  |Y^0|^2-|Y^1|^2-|Y^2|^2=1\ \ {\rm and}\ \ \
  |Y^3|^2+|Y^4|^2+|Y^5|^2=1 \ .
\end{equation}
With the orientation convention advocated in \cite{kl}, $Y^A$'s are
decomposed as twelve `x' and `y' variables as\footnote{Under the
U(1) transformations associated with the positive conserved charges
$M$ and $J_{1,2}$, one has $\delta Y^A= -i Y^A $ for $A=0,1,2$,
while under those associate with $Q_{1,2,3}$, $\delta Y^A=iY^A$ for
$A=3,4,5$.}
\begin{eqnarray}\label{complex}
  && Y^0=X^{-1}-iX^0, \; Y^1=X^1-iX^2, \; Y^2=X^3-iX^5, \nonumber \\
  &&  Y^3(\equiv Z^1)=X^5+iX^6, \;
  Y^4(\equiv Z^2)=X^7+iX^8, \; Y^5(\equiv Z^3)=X^9+iX^{10}  .
\end{eqnarray}
In terms of 12 dimensional Dirac spinors (with $64$ complex
components), the $\frac{1}{16}$ supersymmetry condition is
\begin{equation}\label{susy}
  \Gamma_{A}\Psi=0\ \ (A=1,2,\cdots,6)
\end{equation}
with the above choice (\ref{complex}) of complex structure. The
above 6 projectors define $\frac{1}{16}$ supersymmetry from 10
dimensional IIB viewpoint, since two of the above six are those
reducing $\Psi$ to IIB spinors \cite{kl}. The spinor $\Psi$
satisfying this condition is the Clifford vacuum in 64 dimensional
Hilbert space. $D3$ brane configurations preserving this
supersymmetry should satisfy
\begin{equation}\label{kappa}
  \frac{1-\Gamma^{\hat{r}_1 \hat{r}_2}}{2}(\Gamma-1)\Psi=0\ ,\ \
  \Gamma=-\frac{i}{4!}\epsilon^{\mu\nu\rho\sigma}\gamma_{\mu\nu\rho\sigma}
\end{equation}
where greek indices denote the components in local orthonormal frame
basis of the worldvolume ($\mu=0,1,2,3$). The unit vectors
$\hat{r}_1$ and $\hat{r}_2$ appearing in the superscripts are
normals of $AdS_5$ and $S^5$, respectively. In complex coordinates,
their components are $(Y^0,Y^1,Y^2)$ and $(Y^3,Y^4,Y^5)\equiv
(Z^1,Z^2,Z^3)$. With the condition (\ref{susy}), the supersymmetry
requirement (\ref{kappa}) is simplified as
\begin{eqnarray}
  &&\hspace{-2cm}\frac{1-\Gamma^{\hat{r}_1 \hat{r}_2}}{2}\left[
  -i\epsilon^{\mu\nu\rho\sigma}\left(
  -\frac{1}{2}(t_\mu^{A}t_\nu^{B}
  t_\rho^{\bar{A}}t_\sigma^{\bar{B}})\eta_{A\bar{A}}\eta_{B\bar{B}}+
  \frac{1}{2}(t_\mu^{A}t_\nu^{\bar{B}}
  t_\rho^{\bar{C}}t_\sigma^{\bar{D}})\eta_{A\bar{B}}\Gamma_{\bar{C}\bar{D}}
  \right.\right.\nonumber\\
  &&\left.\left.\hspace{5.5cm}+\frac{1}{24}
  (t_\mu^{\bar{A}}t_\nu^{\bar{B}}t_\rho^{\bar{C}}t_\sigma^{\bar{D}})
  \Gamma_{\bar{A}\bar{B}\bar{C}\bar{D}}\right)-1\right]\Psi=0
  \label{kappa-creation}
\end{eqnarray}
where $t_\mu$'s are orthonormal tangent vectors on the worldvolume,
push-forwarded to the bulk ($A,B=1,\cdots,6$):
\begin{equation}\label{orthogonal}
  \eta_{A\bar{B}}(t_\mu^{A}t_\nu^{\bar{B}}+
  t_\nu^{A}t_\mu^{\bar{B}})=\eta_{\mu\nu}\ .
\end{equation}
The second and third terms in (\ref{kappa-creation}) containing
$\Gamma_{\bar{A}\bar{B}}\Psi$ and
$\Gamma_{\bar{A}\bar{B}\bar{C}\bar{D}}\Psi$ are `excited states'
from the Clifford vacuum, which by themselves cannot cancel the
first and last term proportional to $\Psi$. One should suitably
choose the tangent vectors $t_\mu^A$ to make the second and third
terms to be proportional to $\Psi$. One way is to let these unwanted
terms vanish themselves. Another way would be to use the projector
$\frac{1-\Gamma^{\hat{r}_1\hat{r}_2}}{2}$ in front, i.e., to have
the unwanted terms be annihilated by this projector.

As for the first possibility, we require the worldvolume be given by
a holomorphic hyperspace $\Sigma_6$ with complex dimension 3 in
${\mathbb{C}}^{1+2}\times {\mathbb{C}}^3$. This requirement is met
by a zero locus of three holomorphic functions of $Y^A$.
Intersecting it with $AdS_5\times S^5$ would give the 4 dimensional
worldvolume $\Sigma_{4}$. The tangent space of $\Sigma_6$ is closed
under the action of complex structure, i.e., $I.v\subset
T(\Sigma_6)$ if $v\subset T(\Sigma_6)$. Since we make two
projections on tangent vectors associated with (\ref{constraint}),
two of the four tangent vectors in $T(\Sigma_4)$ should rotate into
$\hat{r}_1$, $\hat{r_2}$ directions by the action of $I$. The
remaining 2-plane is invariant under $I$: following \cite{mikh}, we
call this subspace $T_0(\Sigma_4)\in T(\Sigma_4)$. Note that this
subspace is spacelike since we projected out one of the two timelike
directions in (\ref{constraint}). We choose the two orthonormal
bases of this subspace as $t_z$ and $t_{\bar{z}}=(t_z)^\ast$ in
complex basis, which satisfy
\begin{equation}\label{holomorphicity}
  t_z^{\bar{A}}=0\ .
\end{equation}
Let us also write other two unit vectors normal to this subspace as
$t_a$ ($a=1,2$). The orthonormality condition (\ref{orthogonal})
simplifies as
\begin{eqnarray}
  \eta_{A\bar{B}}t_z^A
  t_{\bar{z}}^{\bar{B}}&=&\eta_{z\bar{z}}=\frac{1}{2}\nonumber\\
  \eta_{A\bar{B}}t_z^A t_{a}^{\bar{B}}&=&0\ \ (a=1,2)\ .
\end{eqnarray}
By requiring holomorphicity (\ref{holomorphicity}), the third term
in (\ref{kappa-creation}) vanishes and one is left with
\begin{eqnarray}\label{kappa-holo}
  0&=&\frac{1-\Gamma^{\hat{r}_1 \hat{r}_2}}{2}\left[
  -i\epsilon^{z\bar{z}a b}\left((\eta_{A\bar{A}}t_z^{A}t_{\bar{z}}^{\bar{A}})
  (\eta_{B\bar{B}}t_a^{B}t_b^{\bar{B}})+\frac{1}{2}
  (\eta_{A\bar{B}}t_z^{A}t_{\bar{z}}^{\bar{B}})
  (t_a^{\bar{C}}t_b^{\bar{D}}\Gamma_{\bar{C}\bar{D}})\right)
  -1\right]\Psi\nonumber\\
  &=&\frac{1-\Gamma^{\hat{r}_1 \hat{r}_2}}{2}\left[-i\epsilon^{z\bar{z}a b}
  \left(\frac{1}{2}(\eta_{A\bar{B}}t_a^{A}t_b^{\bar{B}})+
  \frac{1}{4}(t_a^{\bar{A}}\Gamma_{\bar{A}})
  (t_b^{\bar{B}}\Gamma_{\bar{B}})\right)-1\right]\Psi\ .
\end{eqnarray}

For the last equation to hold, one should impose some nice property
to the remaining two vectors $t_a$. To guess what it can be, note
that the projector $\frac{1-\Gamma^{\hat{r}_1\hat{r}_2}}{2}$ has the
property
\begin{equation}\label{weight projection}
  \frac{1-\Gamma^{\hat{r}_1\hat{r}_2}}{2}
  (\Gamma_{\hat{r}_1}-\Gamma_{\hat{r}_2})=0\ \ \ \ \
  (\Gamma_{\hat{r}_1}=-\Gamma^{\hat{r}_{1}})\ .
\end{equation}
Using (\ref{susy}), the matrix
$\Gamma_{\hat{r}_1}-\Gamma_{\hat{r}_2}$ acted on $\Psi$ is
\begin{equation}\label{projector-anti}
  (\Gamma_{\hat{r}_1}-\Gamma_{\hat{r}_2})\Psi= \left(
  \sum_{A=0,1,2}Y^{\bar{A}}\Gamma_{\bar{A}} - \sum_{A=3,4,5}
  Y^{\bar{A}}\Gamma_{\bar{A}} \right)\Psi\ .
\end{equation}
To use the above properties, we require that the following null
vector
\begin{equation}\label{null tangent}
  t_+^A\equiv\frac{1}{2}(-iY^0,-iY^1,-iY^2,iY^3,iY^4,iY^5)\ ,\ \ \
  t_+^{\bar{A}}=(t_+^A)^\ast
\end{equation}
be a tangent vector. With this vector, one can rewrite
(\ref{projector-anti}) as
\begin{equation}
  (\Gamma_{\hat{r}_1}-\Gamma_{\hat{r}_2})\Psi=
  -2i(t_+^{\bar{A}}\Gamma_{\bar{A}})\Psi\ \ \ ,\ \ \
  (\Gamma_{\hat{r}_1}-\Gamma_{\hat{r}_2})=
  +2i(t_+^{A}\Gamma_{A})-2i(t_+^{\bar{A}}\Gamma_{\bar{A}})\ .
\end{equation}
We also write the last tangent vector with the symbol $t_-$, which
we also choose to be null. The two vectors are decomposed into a
transverse time-evolution vector and a space-like tangent vector. We
write $t_\pm=\gamma_{\pm}({\bf e}_\psi\pm {\bf e}_\phi)$ with
\begin{equation}\label{decompose null}
  {\bf e}_\phi=\cosh\rho\ (e_t, ve_\phi)\ \ ,\ \ \
  {\bf e}_\psi=\cosh\rho\ \sqrt{1-v^2}(0,e_\psi)\ \ ,
\end{equation}
where $e_\phi$ and $e_\psi$ are unit vectors in
${\mathbb{R}}^{10}\subset{\mathbb{R}}^{2+10}$ normal/tangent to the
spatial configuration at given time, respectively: they are mutually
orthogonal $e_\phi\cdot e_\psi=0$. $e_t$ is the unit time vector in
${\mathbb{R}}^{2}\subset{\mathbb{R}}^{2+10}$, and $v$ is the
physical (=transverse) velocity. $\gamma_{\pm}$ are boost
ambiguities in choosing the null local orthonormal frames preserving
the canonical form of the metric
\begin{equation}\label{orthonormal metric}
  \eta_{+-}=\eta_{z\bar{z}}=\frac{1}{2}\ .
\end{equation}
$\gamma_\pm$ are thus required to satisfy
\begin{equation}\label{lightlike}
  \eta_{+-}=\frac{1}{2}=\eta_{A\bar{B}}(t_+^At_-^{\bar{B}}+t_-^At_+^{\bar{B}})=
  2(1-v^2)\gamma_+\gamma_- \cosh^2\rho \ .
\end{equation}
Since our choice (\ref{null tangent}) already fixed
$\gamma_+=\frac{1}{2}$, we have
\begin{equation}
  t_+=\frac{1}{2}({\bf e}_\psi+{\bf e}_\phi)\ ,\ \
  t_-=\frac{1}{2(1-v^2)\cosh^2\rho}({\bf e}_\psi-{\bf e}_\phi)\ .
\end{equation}
Futhermore, the the normalization (\ref{orthonormal metric})
requires
\begin{equation}\label{epsilon}
  \epsilon^{z\bar{z}+-}\
  (=(-2i)\cdot(-2)\epsilon^{xy\psi t})=+4i\ \
  (\epsilon_{0123}=1)
\end{equation}
in the orientation convention of \cite{kl}. We note that
\begin{equation}\label{light norm}
  \eta_{A\bar{B}}(t_+^A t_-^{\bar{B}}-t_-^A t_+^{\bar{B}})\sim
  (t_-,I.t_+)\sim(t_-,\hat{r}_1-\hat{r}_2)=0\ \ \rightarrow\ \
  \eta_{A\bar{B}}t_+^A t_-^{\bar{B}}=\eta_{A\bar{B}}t_-^A t_+^{\bar{B}}
  =\frac{1}{4}\ ,
\end{equation}
since $t_-$ is orthogonal to both normal vectors $\hat{r}_1$ and
$\hat{r}_2$. Having fixed all the tangent vectors, we completely
specified our ansatz to solve the BPS equation.

With the above ansatz, the supersymmetry condition
(\ref{kappa-holo}) becomes
\begin{eqnarray}\label{susy final}
  0&=&\frac{1-\Gamma^{\hat{r}_1 \hat{r}_2}}{2}\left[-i\epsilon^{z\bar{z}+-}
  \left\{\frac{1}{2}\eta_{A\bar{B}}(t_+^{A}t_-^{\bar{B}}-t_-^{A}t_+^{\bar{B}})-
  \frac{i}{4}(t_-^{\bar{A}}\Gamma_{\bar{A}})
  (\Gamma_{\hat{r}_1}-\Gamma_{\hat{r}_2})\right\}-1\right]\Psi\nonumber\\
  &=&\frac{1-\Gamma^{\hat{r}_1 \hat{r}_2}}{2}
  \left[-i\epsilon^{z\bar{z}+-}(\eta_{A\bar{B}}t_+^{A}t_-^{\bar{B}})
  -1\right]\Psi=0
\end{eqnarray}
using (\ref{epsilon}), (\ref{light norm}). This completes the proof
that any holomorphic hyperspace which includes
$i(-Y^0,-Y^1,-Y^2,Y^3,Y^4,Y^5)$ as a tangent vector defines a
worldvolume of $\frac{1}{16}$-BPS $D3$-brane. To summarize, the
solution is given by three holomorphic `weighted-homogeneous'
functions
\begin{eqnarray}\label{scalar solution}
  &&F_k(Y^0,Y^1,Y^2,Z^1,Z^2,Z^3)=0\ \ \ \ (k=1,2,3)\ , \nonumber\\
  &&{\rm where }\ \ \ F_k(\lambda Y^A,  \lambda^{-1}Z^A)=\lambda^{d_k}
  F_k(Y^A,Z^A)\ .
\end{eqnarray}
Note that $Z^A\!=\!Y^{A+2}$ for $A\!=\!1,2,3$. The numbers $d_k$ are
suitable degrees of $F_k$. We stress that the $\pm$ weights of the
coordinates $Z^A$ and $Y^A$ ($A=1,2,3$) is a consequence of
(\ref{weight projection}), which in turn required (\ref{null
tangent}) to be a tangent vector with the relative $-$ sign.

Now we show that the energy of this configuration saturates the BPS
bound given by the sum of five $U(1)^5\in SO(6)\times SO(4)$ angular
momenta. The energy conjugate to the time $t$ (appearing in
$Y^0=\cosh\rho\ e^{-it}$) comes from two contributions. That coming
from the DBI action reads
\begin{equation}\label{dbi energy}
  M_{DBI}=N\int_{\Sigma_3}\cosh\rho\ vol(\Sigma_3)\frac{1}{\sqrt{1-v^2}}
\end{equation}
where $\Sigma_3$ denotes the spatial configuration of the brane in
$AdS_5\times S^5$ at given time, and $N$ is the 5-form flux number.
The volume of $\Sigma_3$ in the integrand is measured by the
pull-back of the bulk metric. The energy also gets contribution from
the Wess Zumino term if the brane is extended in $AdS_5$:
\begin{equation}
  M_{WZ}=-N\int_{\partial^{-1}\Sigma_3} 4i(e^t).i({e}^{\hat{r}_1}).\
  \frac{1}{6}\tilde\omega\wedge\tilde\omega\wedge\tilde\omega
\end{equation}
where
\begin{equation}
  \omega=\frac{i}{2}\sum_{A=1}^3
  \eta_{A\bar{A}}dZ^A\wedge d\bar{Z}^{\bar{A}}\ \ {\rm and}\ \ \
  \tilde\omega=\frac{i}{2}\sum_{A,B=0}^2\eta_{A\bar{B}}dY^A\wedge
  d\bar{Y}^B
\end{equation}
are the K\"{a}hler forms of ${\mathbb{C}}^3$ and
${\mathbb{C}}^{2+1}$ respectively, $e^{\hat{r}_2}\sim Z^A$ and
$e^{\hat{r}_1}\sim Y^A$ (in components) are vectors normal to $S^5$
and $AdS_5$, respectively, and $e^t=(-iY^0,0,0)$ is the time
translation vector. The sum of five angular momenta from the DBI
action is
\begin{equation}\label{dbi charge}
  [Q_1+Q_2+Q_3+J_1+J_2]_{DBI}=N\int_{\Sigma_3}\cosh\rho\ vol(\Sigma_3)
  \frac{v^2}{\sqrt{1-v^2}}\ ,
\end{equation}
which comes from the fact that the diagonal of $U(1)^5$ rotations is
generated by the vector $\dot{\vec{X}}=(iZ^A,-iY^1,-iY^2)$,
decomposed to transverse part $\cosh\rho\ \vec{v}$ and longitudinal
one $\cosh\rho\ \sqrt{1-v^2}\ e_\psi$. The `sum of momenta minus the
energy' from the Wess-Zumino term is computed in the same way as
\cite{mikh}:
\begin{eqnarray}\label{wz charge-energy}
  \hspace*{-0.7cm}
  \left.\sum_{A=1^3}Q_A+J_1+J_2-M\right|_{WZ}\hspace{-0.3cm}
  &=&\frac{iN}{2}\int_{\Sigma_3}(Z\cdot d\bar{Z}-\bar{Z}\cdot dZ)
  \wedge\omega-(Y\cdot d\bar{Y}-\bar{Y}\cdot dY)\wedge\tilde\omega\nonumber\\
  \hspace*{-0.5cm}&=&N\int_{\Sigma_3} e^{\parallel}\wedge \omega
  -\tilde{e}^{\parallel}\wedge\tilde\omega\equiv
  N\int_{\Sigma_3} I.e^{\hat{r}_2}\wedge \omega-
  I.e^{\hat{r}_1}\wedge\tilde\omega
\end{eqnarray}
where $(I.e^{\hat{r}_2},I.e^{\hat{r}_1})$ in this integral denotes
the 1-form dual to the real 10 dimensional vector
$(iZ^A,iY^1,iY^2)$, pull-backed to the world-volume. Note that the
sign of the second term in the last integrand can be checked
independently by requiring it be positive for $\frac{1}{2}$-BPS dual
giant gravitons, for which $(0,-I.e^{\hat{r}_1})$ is the tangent
vector. For the BPS energy to be given by the sum of five charges
$Q_A$ and $J_{1,2}$, we want (\ref{dbi energy}) $=$ (\ref{dbi
charge}) $+$ (\ref{wz charge-energy}), i.e.,
\begin{equation}
  \int_{\Sigma_3} e^{\parallel}\wedge \omega
  -\tilde{e}^{\parallel}\wedge\tilde\omega\stackrel{!}{=}
  (\ref{dbi energy})-(\ref{dbi charge})=
  \int_{\Sigma_3}\cosh\rho\ \sqrt{1-v^2}\ vol(\Sigma_3)
  \equiv\int_{\Sigma_3}(e^{\parallel}-\tilde{e}^{\parallel})
  \wedge(\omega+\tilde{\omega})\ .
\end{equation}
$e^{\parallel}-\tilde{e}^{\parallel}$ in the last integral is the
1-form dual to $t_+$ in (\ref{null tangent}). To see if this
relation holds, one only needs to check
\begin{equation}\label{cross}
  \int_{\Sigma_3}e^{\parallel}\wedge \tilde{\omega}
  -\tilde{e}^{\parallel}\wedge\omega =0\ .
\end{equation}
Since the integrand of (\ref{cross}) is exact,
\begin{equation}
  e^{\parallel}\wedge \tilde{\omega}
  -\tilde{e}^{\parallel}\wedge\omega\sim
  d\left(e^{\parallel}\wedge\tilde{e}^{\parallel}\right)\ ,
\end{equation}
the integral (\ref{cross}) is zero for \textit{compact} D3 branes.
Therfore, the energy is given as
\begin{equation}
  M=Q_1+Q_2+Q_3+J_1+J_2\ ,
\end{equation}
which obeys the same BPS relation as the black hole mass (\ref{AD
mass}), taking into account that the latter is conjugate to $\ell t$
in our coordinate.

\section{Examples of giant gravitons}

The geometry of the classical giant graviton solutions, given by
holomorphic functions, is expected to be rich. One class which is
relatively easy to examine is those whose world-volumes are nearly
degenerate. A well-known degeneration is the spherical giant
graviton shrinking to a point particle as angular momentum
decreases. The $\frac{1}{2}$-BPS configuration
\begin{equation}
  Z^3 Y^0=\alpha \ ,\ \ Y^1=Y^2=0
\end{equation}
degenerates to a point as $|\alpha|$ increases to $1$.

In this section, as simple examples of the giant gravitons with
nontrivial topology, we investigate various ways the $D3$ branes can
shrink, where the 3+1 dimensional worldvolumes degenerate to 1+1
dimensional or 2+1 dimensional objects.\footnote{For
$\frac{1}{8}$-BPS giant gravitons, there is another limit where the
geometry is easy to analyze. It is the stationary giant gravitons in
$S^5$ given by a single homogeneous function. They are given by
circle fibrations over complex algebraic curves.} The
nearly-degenerate configurations are easy to analyze, and we show
that the world-volumes of the $D3$ branes we treat in this section
are topologically tubular: $S^1\times S^2$ or $S^1\times S^1\times
S^1$. The cross sections $S^2$ or $S^1\times S^1$ of the tubes are
small in our examples, just like small $S^3$ for nearly point-like
gravitons. One obvious interpretation of these objects would be
superpositions of point-like gravitons, which may also be supported
by \cite{bglm} in the $\frac{1}{8}$-BPS sector. We would also like
to give them an interpretation in terms of tensionless strings.

\subsection{$\frac{1}{4}$- and $\frac{1}{8}$-BPS examples}

We first investigate the giant gravitons in $\frac{1}{4}$- and
$\frac{1}{8}$-BPS sectors, which were treated by Mikhailov. For
simplicity, let us first consider the following $\frac{1}{4}$-BPS
solution
\begin{equation}\label{quarter string}
  Z^1 Z^2 (Y^0)^2 =\alpha\ \ ,\ \ \ Y^1=Y^2=0\ .
\end{equation}
This actually belongs to a class of $\frac{1}{8}$-BPS solutions
discussed in \cite{mikh},
\begin{equation}\label{1/8 membrane}
  (Z^1)^{m_1}(Z^2)^{m_2}(Z^3)^{m_3} (Y^0)^{m_1+m_2+m_3}=\alpha\ ,
  \ \ Y^1=Y^2=0\
\end{equation}
with integral $m_A$. The three $S^5$ momenta $Q_A$ are proportional
to the vector $(m_1,m_2,m_3)$, so (\ref{quarter string}) carries
$Q_1=Q_2$ and $Q_3=0$. We would like to study the geometry of this
brane as one increases $\alpha$, which one may simply choose to be a
positive number. Since we are interested in the intersection of
(\ref{quarter string}) with $S^5$, the variables $Z^1$ and $Z^2$ are
bounded from above. This means that the configuration given by the
equation (\ref{quarter string}) will not intersect $S^5$ any more if
$\alpha$ is too large. The maximal value to have nonempty
intersection is $\alpha=\frac{1}{2}$, for which one obtains
\begin{equation}
  Z^3=0\ ,\ \ Z^1=\frac{1}{\sqrt{2}}\ e^{i\phi}\ ,\ \
  Z^2=\frac{1}{\sqrt{2}}\ e^{-i\phi}\ .
\end{equation}
As for the phases, the equation (\ref{quarter string}) requires
$\phi_1\!=\!-\phi_2$($\equiv\phi$). One can see that this
configuration degenerates to a circular string in $S^5$
parameterized by the angle $\phi$, located at
$|Z^1|=|Z^2|=\frac{1}{\sqrt{2}}$ (and $Z^3=0$). To see how the $D3$
brane degenerates, one may check a slightly resolved configuration
$\alpha=\frac{1}{2}-\epsilon^2$, where $\epsilon\ll 1$ is a small
positive number. With the following parametrization
\begin{equation}\label{s5 coordinate}
  Z^1=\sin\Theta\cos\theta\ e^{i\phi^1}\ ,\ \
  Z^2=\sin\Theta\sin\theta\ e^{i\phi^2}\ ,\ \
  Z^3=\cos\Theta\ e^{i\phi^3}
\end{equation}
of $S^5$, one obtains
\begin{equation}
  \frac{1}{2}\left(\Theta-\frac{\pi}{2}\right)^2+
  \left(\theta-\frac{\pi}{4}\right)^2\approx\epsilon^2\ ,\ \
  \phi^1=-\phi^2\ (\equiv\phi)\ .
\end{equation}
The angle $\phi^3$ is a tangent direction of the brane. The
variables $\Theta$, $\phi^3$ and $\theta$ form an ellipsoid whose
size is $\epsilon$, while the angle $\phi$ parameterizes the
macroscopic circle of the string. Therefore, the configuration is
topologically a thin tube with $S^2$ cross section.

The time evolution of this nearly degenerate `string' is given by
$\dot{\phi}^1=\dot{\phi}^2=1$, with a lightlike transverse velocity.
Since the motion is nearly lightlike for small $\epsilon$, the
energy and the angular momenta are dominantly given by the DBI
action for nearly-degenerated branes. To the leading order in
$\epsilon$,  they are given as
\begin{equation}
  Q_1=Q_2\approx\frac{N}{4\pi^2}\int_{S^1\times S^2}
  \frac{1}{\sqrt{1-v^2}}\approx 2N\epsilon
\end{equation}
while $Q_3=0$ exactly \cite{mikh}. We used the following expression
for the transverse speed,
\begin{equation}
  v^2=\frac{4\alpha^2}{|Z^1|^2+|Z^2|^2}=(1-2\epsilon^2)\sin(2\theta)
  \ \ \rightarrow\ \ \ 1-v^2\approx 2\epsilon^2+
  2\left(\theta-\frac{\pi}{4}\right)^2
\end{equation}
up to $O(\epsilon)^4$ corrections that we can ignore.

As mentioned in the beginning of this section, the above
configuration may be interpreted as a collection of nearly
point-like giant gravitons: at least in the degenerate limit, it
looks like a superposition of super-gravitons. The way this object
expands in the above solution, as angular momenta $Q_1=Q_2$
increase, is a single tube with $S^2$ cross section. The above
configuration has been also treated in the literature as a
tensionless string solution \cite{mmt2}. The analysis of the
tensionless strings is reviewed in the appendix, generalizing
\cite{mmt2} to the $\frac{1}{16}$-BPS sector. The shape of the
$\frac{1}{4}$-BPS solution with $Q_1=Q_2$ agrees with what we
presented above with giant gravitons.\footnote{Originally, these
tensionless strings were regarded as an ultra-relativistic limit of
the nearly-BPS Nambu-Goto strings in $AdS_5\times S^5$
\cite{mmt1,mmt2}. The above giant graviton is not directly related
to the latter. As a BPS cousin of the fundamental strings, one may
put (local) fundamental string charges to the above tubular brane by
turning on electric flux along the string direction, following
\cite{kl}.}

$\frac{1}{4}$-BPS tensionless strings with $Q_1\neq Q_2$ can also be
reproduced from the degeneration of giant gravitons. It is simply
given by the equation
\begin{equation}
  (Z^1)^{m_1}(Z^2)^{m_2}=\alpha\ \ \ {\rm with}\ \ \
  \alpha\rightarrow\sqrt{\frac{(m_1)^{m_1}(m_2)^{m_2}}
  {(m_1+m_2)^{m_1+m_2}}}\ .
\end{equation}
The ratio of charges are given as $(Q_1,Q_2)\propto (m_1,m_2)$,
whose value is determined by the $\alpha$ parameter as in the
previous example. The string is located in the limit at
\begin{equation}
  Z^3=0\ ,\ \ |Z^1|=\sqrt{\frac{m_1}{m_1+m_2}}\ ,\ \
  |Z^2|=\sqrt{\frac{m_2}{m_1+m_2}}
\end{equation}
while satisfying $m_1\phi^1+m_2\phi^2=0$. The string is
parameterized by an angle $\sigma$ with range $0\sim 2\pi$, which is
related to $\phi^{1,2}$ by $\phi^1=m_2\sigma$, $\phi^2=-m_1\sigma$.
The shape and the charges again agree with the tensionless string
solutions. From the result derived in our appendix, the agreement is
clear if one sets $Q_3=J_1=J_2=0$.

One can also investigate (nearly) degenerate configurations in the
$\frac{1}{8}$-BPS sector with (\ref{1/8 membrane}). Near the maximal
value $\alpha=\sqrt{\frac{(m_1)^{m_1}(m_2)^{m_2}(m_3)^{m_3}}
{(m_1+m_2+m_3)^{m_1+m_2+m_3}}}$, the topology of the $D3$ brane is
$S^1\times S^1\times S^1$ where the last $S^1$ shrinks. It is
located at
\begin{equation}\label{1/8 radii}
  |Z^1|=\sqrt{\frac{m_1}{m_1+m_2+m_3}}\ ,\ \
  |Z^2|=\sqrt{\frac{m_2}{m_1+m_2+m_3}}\ ,\ \
  |Z^3|=\sqrt{\frac{m_3}{m_1+m_2+m_3}}
\end{equation}
while macroscopic $S^1\times S^1$ is spanned by three angles
$\phi^A$ subject to a constraint
\begin{equation}\label{1/8 angle}
  m_1\phi^1+m_2\phi^2+m_3\phi^3=0\ .
\end{equation}
In comparison, the tensionless strings in \cite{mmt2} with charges
$Q_A\propto m_A$ are also located at (\ref{1/8 radii}) and stretched
along a line in the above $S^1\times S^1$ described by (\ref{1/8
angle}). The $S^1\times S^1$ ($\times$ small $S^1$) may be
interpreted as a collection of such tensionless strings.

\subsection{$\frac{1}{16}$-BPS examples}

Finally we present similar $\frac{1}{16}$-BPS degenerate objects. We
will also see that the $\pm$ weights in (\ref{scalar solution}) play
some roles for such configurations to exist. For simplicity we only
consider $\frac{1}{16}$-BPS examples with $Q_3=0$.\footnote{General
$\frac{1}{16}$-BPS examples we find are similar to the
$\frac{1}{8}$-BPS ones, degenerating to $2+1$ dimensions.} We take
the following holomorphic functions:
\begin{eqnarray}\label{1/16 degeneration}
  \alpha&=&(Z^1)^{m_1}(Z^2)^{m_2}(Y^0)^{m_1+m_2}\
  ,\nonumber\\
  \frac{Y^1}{Y^0}&=&
  \beta_1 (Z^1)^{p_1}(Z^2)^{p_2}(Y^0)^{p_1+p_2}\\
  \frac{Y^2}{Y^0}&=&
  \beta_2 (Z^1)^{q_1}(Z^2)^{q_2}(Y^0)^{q_1+q_2}\ .\nonumber
\end{eqnarray}
$\vec{p}$ and $\vec{q}$ are rational, while $\vec{m}$ is integral.
We use the the following coordinates of $AdS_5$
\begin{equation}
  Y^0=\cosh\rho\ e^{-it}\ ,\ \
  Y^1=\sinh\rho\ \cos\vartheta\ e^{-i\varphi^1}\ ,\ \
  Y^2=\sinh\rho\ \sin\vartheta\ e^{-i\varphi^2}\ ,
\end{equation}
while the $S^5$ coordinates are same as (\ref{s5 coordinate}). At
$t\!=\!0$, the four angles $\phi^{1,2}$ and $\varphi^{1,2}$ are
constrained as
\begin{equation}\label{string angle 1}
  \vec{m}\cdot\vec{\phi}=0\ \ ,\ \ \
  \varphi^1=-\vec{p}\cdot\vec{\phi}\ ,\ \ \
  \varphi^2=-\vec{q}\cdot\vec{\phi}\ .
\end{equation}
Introducing an angle $\sigma$ (with range $0\!\sim\!2\pi$), one can
write the above four angles as
\begin{eqnarray}\label{string angle 2}
  (\phi^1,\phi^2)&=&(t+m_2 \sigma,\ t-m_1\sigma)\equiv
  t(1,1)+\vec{r}\sigma\nonumber\\
  (\varphi^1,\varphi^2)&=&\left(t+(m_1p_2\!-\!m_2p_1)\sigma,
  \ t+(m_1q_2\!-\!m_2q_1)\sigma\frac{}{}\right)\equiv
  t(1,1)+\vec{s}\sigma
\end{eqnarray}
for general $t$. Since the angle $\phi^3$ is free, there are two
independent angle-like directions tangent to the brane. One
combination of other four variables ($\Theta$, $\theta$, $\rho$,
$\varphi$) parameterizing $|Z^A|$ and $|Y^A|$ forms the last tangent
direction. We first show that the brane can degenerate along this
last direction. The degeneration appears at $|Z^3|=0$, which makes
the $\phi^3$ direction degenerate also. The resulting configuration
would therefore be string-like, which will be identified with the
$\frac{1}{16}$-BPS tensionless strings treated in the appendix.

On the holomorphic hyperspace $\Sigma_6$ in ${\mathbb{C}}^{1+5}$
given by (\ref{1/16 degeneration}), three of the six tangent vectors
lie in the 6 dimensional subspace ${\mathbb{R}}^{1+5}$ of
${\mathbb{C}}^{1+5}$ generated by $|Z^A|$ and $|Y^A|$: by
intersecting it with $AdS_5\times S^5$, one obtains a 1 dimensional
locus which we claim to degenerate. The three normals (1-forms) to
$\Sigma_6$ in ${\mathbb{R}}^{5+1}$, in ($|Z^A|$, $|Y^A|$)
components, are
\begin{eqnarray}
  {\bf a}&\equiv&\left(\frac{m_1}{|Z^1|},\ \frac{m_2}{|Z^2|},\ 0,\
  \frac{m_1+m_2}{|Y^0|},\ 0,\ 0\right)\nonumber\\
  {\bf b}&\equiv&\left(\frac{p_1}{|Z^1|},\ \frac{p_2}{|Z^2|},\ 0,\
  \frac{p_1+p_2+1}{|Y^0|},\ -\frac{1}{|Y^1|},\ 0\right)
  \label{normals}\\
  {\bf c}&\equiv&\left(\frac{q_1}{|Z^1|},\ \frac{q_2}{|Z^2|},\ 0,\
  \frac{q_1+q_2+1}{|Y^0|},\ 0,\ -\frac{1}{|Y^2|}\right)\ .
  \nonumber
\end{eqnarray}
These are gradients of (\ref{1/16 degeneration}). Intersecting the
hyperspace with $AdS_5\times S^5$, the single tangent vector of
$\Sigma_4$ in ${\mathbb{R}}^{5+1}$ is orthogonal to the normals
(\ref{normals}) as well as
\begin{equation}\label{AdS S normal}
  e^{\hat{r}_2}\equiv(|Z^1|,|Z^2|,|Z^3|,0,0,0)\ \ \ {\rm and}\ \ \
  e^{\hat{r}_1}\equiv(0,0,0,|Y^0|,-|Y^1|,-|Y^2|)\ ,
\end{equation}
the normals of $AdS_5\times S^5$. When the (claimed) degeneration
appears, the rank of these five normals should reduce so that the
`tangent' vector orthogonal to them would be ambiguous. In other
words, one has a linear relation between the above five normals,
\begin{equation}\label{relation}
  {\bf a}+\lambda_1{\bf b}+\lambda_2{\bf c}=
  \mu e^{\hat{r}_2}+\nu e^{\hat{r}_1}
\end{equation}
with suitable coefficients. We rewrite them in components as
\begin{eqnarray}
  \hspace*{-1.5cm}1{\rm st\ and}\ 2{\rm nd}&:&
  \left(\ |Z^1|^2,\ |Z^2|^2\ \right)=
  \frac{\vec{m}+\lambda_1\vec{p}+\lambda_2\vec{q}}{\mu}\label{sol 1}\\
  \hspace*{-1.5cm}3{\rm rd}&:&|Z^3|=0\label{sol 2}\\
  \hspace*{-1.5cm}4{\rm th}&:&1=
  \frac{m_1+m_2}{\nu}+(p_1+p_2)|Y^1|^2+(q_1+q_2)|Y^2|^2
  \label{sol 3}\\
  \hspace*{-1.5cm}5{\rm th\ and}\ 6{\rm th}&:&
  \left(\ |Y^1|^2,\ |Y^2|^2\ \right)=
  \frac{\vec{\lambda}}{\nu}\label{sol 4}\ .
\end{eqnarray}
Note that, applying the requirement $|Z^1|^2+|Z^2|^2+|Z^3|^2=1$ to
(\ref{sol 1}) and (\ref{sol 2}), one obtains
\begin{equation}
  \frac{\mu}{\nu}=\frac{m_1+m_2}{\nu}+(p_1+p_2)|Y^1|^2+
  (q_1+q_2)|Y^2|^2\ .
\end{equation}
Comparing it with (\ref{sol 3}), one obtains $\mu=\nu$. Here, the
fact that the weights of $Z^A$ and $Y^A$ are opposite in counting
homogeneity of the holomorphic functions (\ref{scalar solution}) is
crucial for the degenerate configurations to exist.\footnote{For
instance, consider a hyperspace given by genuine homogeneous
functions, where $Y^0$ factors in the right hand sides of (\ref{1/16
degeneration}) are inverted. Then the terms with $|Y^1|^2$ and
$|Y^2|^2$ in (\ref{sol 3}) would acquire additional $-$'s,
incompatible with (\ref{sol 1}) unless $|Y^a|^2=0$.}

The charges of the nearly degenerated branes are again dominantly
given by DBI contributions,
\begin{equation}
  Q_A\approx\int_{\Sigma_3} vol(\Sigma_3)
  \frac{|Z^A|^2}{\sqrt{1-v^2}}\ ,\ \
  J_a\approx\int_{\Sigma_3} vol(\Sigma_3)
  \frac{|Y^a|^2}{\sqrt{1-v^2}}
\end{equation}
where $v^2\approx 1$ ($A=1,2,3$, $a=1,2$). For the above
configurations, where $|Z^A|$ and $|Y^a|$ are nearly constants, the
charges are proportional to
\begin{equation}
  (\vec{Q},\ \vec{J})\propto(|Z^A|^2,\ |Y^a|^2)\ .
\end{equation}
From this structure, one can reproduce (\ref{string locus 1}),
(\ref{string locus 2}) and (\ref{string locus 3}) for the
tensionless strings. Also, note that the charges ($\vec{Q}$,
$\vec{J}$) are orthogonal to the tangent direction generated by
$\phi^3$ and $\sigma$ translations in (\ref{string angle 2}). The
argument is the same as that in \cite{mikh} for (\ref{1/8
membrane}). As the result, one obtains
\begin{equation}
  \vec{r}\cdot\vec{Q}+\vec{s}\cdot\vec{J}=0\ ,\ \ Q_3=0\ ,
\end{equation}
which is just (\ref{string constraint}) in the $Q_3\!=\!0$ case.

\section{Concluding remarks}

In this paper we explored the $\frac{1}{16}$-BPS objects of type IIB
string theory in $AdS_5\times S^5$. First, in the gravity side, we
investigated the supersymmetric black holes in gauged supergravity
and pointed out that there is a simple expression for the
Bekinstein-Hawking entropy in terms of physical charges. We think
the simplicity of the expression suggests the existence of its nice
microscopic explanation. In the microscopic side, we investigated
the classical solutions of $\frac{1}{16}$-BPS giant gravitons. We
obtained a class of solutions given by three holomorphic functions
which are homogeneous with suitable weights on the bulk coordinates.
As examples, we studied string-like degenerations of these
solutions.

An obvious work to be done in the microscopic side would be to see
whether there are more general $\frac{1}{16}$-BPS solutions than
(\ref{scalar solution}). Also, one may hope to obtain clues of
understanding the black hole microphysics from the solution
(\ref{scalar solution}) or its generalizations. Especially, it will
be very interesting to see whether there are microscopic
configurations saturating the angular momentum bound (\ref{angular
bound}). An analogous issue has been addressed for the rotating
objects in Minkowski space. For instance, the entropy of rotating
$D1$-$D5$ on $T^5$ (or $K3$) is $S=2\sqrt{2}\pi\sqrt{Q_1Q_5-J}$ (or
$4\pi\sqrt{Q_1Q_5-J}$). Here the zero entropy configurations are
given by the circular supertubes \cite{ma-to,pa-mar,bho,bhko}, or
U-duals of them \cite{dghw,lu-ma,ch-oh,mnt}. With the $AdS_5$ black
holes, it may be interesting to consider the 2-charge black holes
with $S=2\pi\sqrt{Q_1Q_2-N^2 J}$, even if the $N^2$ factor makes the
problem more challenging.

One interesting feature of the $\frac{1}{16}$-BPS solutions
discovered in this paper is that, they may be viewed as
$\frac{1}{16}$-BPS deformations of either $\frac{1}{8}$-BPS giant
gravitons of \cite{mikh}, or the $\frac{1}{8}$-BPS dual giant
gravitons of \cite{ma-sur}. For instance, the three holomorphic
equations can be arranged in one of the two forms
\begin{eqnarray}
  &&F(Z^AY^0)=0\ ,\ \ \frac{Y^a}{Y^0}=f^a(Z^AY^0)\ \ \
  (a=1,2)\ \ \label{giant}\\
  &&Z^AY^0=F^A\left(\frac{Y^1}{Y^0},\frac{Y^2}{Y^0}\right)\ \
  (A=1,2,3)\ ,\label{dual}
\end{eqnarray}
at least locally. (\ref{giant}) is a deformation of \cite{mikh}, the
latter being given by $f^a\equiv 0$. (\ref{dual}) may be viewed as a
deformation of dual giant gravitons discussed in \cite{ma-sur},
given by $F^A=c^A$ where $c^A$'s are constants. It will be
interesting to further explore these dual descriptions, since they
seem to be amalgamated into a single solution space  in the
$\frac{1}{16}$-BPS sector.

As for our work on $AdS_5$ black holes, we stress that the
Bekenstein-Hawking entropy of $AdS_5$ black holes admit simple
expressions not only in the $U(1)^3$ supergravity (truncating
$SO(6)$ gauged supergravity), obtained by an $S^5$ reduction of type
IIB string theory. The only requirement for the supergravity is that
the scalars in the vector multiplet live on a symmetric space. For
instance, this is accomplished by $AdS_5$ supergravity with no less
than $16$ supersymmetries \cite{aw-to}. For instance, one can obtain
${\mathcal{N}}=2$ $AdS_5$ supergravities by compactifying M-theory
on suitable 6-manifolds. An explicit example, known as the
${\mathcal{N}}=2$ Maldacena-Nunez solution \cite{ma-nu}, falls into
this class. It will be interesting to consider black holes in this
background, carrying electric charges coming from the membrane
wrapping an internal 2-cycle, as well as those coming from internal
momenta (analogous to the $S^5$ Cartans). In this case, the central
charge factor $c$ in front of the angular momentum $J$ is
proportional to $N^3$, where $N$ is the number of $M5$ branes on
which the dual superconformal field theory lives.

\vskip 0.5cm

\hspace*{-0.6cm}{\bf\large Acknowledgements}

\hspace*{-0.6cm}We are grateful to Sangmin Lee for many useful
discussions and suggestions, and especially for checking some
statements in this paper independently. We would also like to thank
Juan Maldacena, Gautam Mandal, Shiraz Minwalla, Soo-Jong Rey and
Piljin Yi for discussions and comments at various stages of this
work. This work is  supported in part by  KOSEF Grant
R010-2003-000-10391-0 (S.K, K.L.), KOSEF SRC Program through CQUeST
at Sogang University (K.L.), KRF Grant No. KRF-2005-070-C00030
(K.L.).  One of us (K.L.) appreciates Perimeter Institute where the
work is done partially.

\begin{appendix}

\section{The $\frac{1}{16}$-BPS tensionless strings}

In this appendix we obtain the tensionless string solutions that are
$\frac{1}{16}$-BPS, which have BPS energy $M=Q_1+Q_2+Q_3+J_1+J_2$.
The bosinic action for the tensionless string is, generalizing that
of \cite{mmt2},
\begin{eqnarray}\label{tension action}
  {\mathcal{L}}&=&p_\rho\dot{\rho}+p_{\vartheta}\dot\vartheta+
  \vec{J}\cdot\dot{\vec{\varphi}}+
  p_\theta\dot\theta+p_{\psi}\dot\psi+
  +\vec{Q}\cdot\dot{\vec{\phi}}\nonumber\\
  &&-\cosh\rho\sqrt{p_\rho^{\ 2}+
  \frac{p_{\vartheta}^{\ 2}}{\sinh^2\rho}+
  \frac{1}{\sinh^2\rho}\left(\frac{J_1^{\ 2}}{\cos^2\vartheta}+
  \frac{J_2^{\ 2}}{\sin^2\vartheta}\right)+
  p_\theta^{\ 2}+\frac{p_\psi^{\ 2}}{\sin^2\theta}+
  \sum_{A=1}^3\frac{Q_A^{\ 2}}{n_A^{\ 2}}}\nonumber\\
  &&-\lambda\left(p_\rho \rho^\prime+p_{\vartheta}\vartheta^\prime+
  \vec{J}\cdot{\vec{\varphi}}^\prime+
  p_\theta\theta^\prime+p_{\psi}\psi^\prime+
  +\vec{Q}\cdot{\vec{\phi}}^\prime\right)
\end{eqnarray}
where $\rho$, $\vartheta$ and $\varphi_{1,2}$ parameterize the
spatial directions of global $AdS_5$, and $\Theta$, $\theta$, $\psi$
and $\phi^{1,2,3}$ parameterize the $S^5$ with the following metric
\begin{equation}
  ds_{S^5}^{\ 2}=d\theta^2+\sin^2\theta d\psi^2+
  \sum_{A=1}^3 n_A^{\ 2}(d\phi^A)^2\ .
\end{equation}
The coefficients $n_A$ parameterize a unit $S^2$:
$(n_A)=(\cos\theta,\sin\theta\cos\psi,\sin\theta\sin\psi)$, which
are $|Z^A|^2$ in our previous notation. The term in the second line
of (\ref{tension action}) is minus the Hamoiltonian $\mathcal{H}$,
and the last line with Lagrange multiplier $\lambda$ is inserted for
the string reparametrization constraint. The prime denotes the
differentiation with the spatial coordinate of the worldsheet, say
$\sigma$($\sim\sigma+2\pi$).

Generalizing \cite{mmt2}, the solution we are interested in has
constant $\rho$, $\vartheta$ in $AdS_5$ and constant $n_A$'s in
$S^5$, together with $p_\rho=p_\vartheta=p_\theta=p_\psi=0$. Other
five momena $\vec{Q}$, $\vec{J}$ as well as the energy $\mathcal{H}$
are constants of motion when integrated with $\sigma$. We are
interested in determining the values of these constant coordinates
in terms of the constants of motion.

We start from the equation of motion coming from $\delta n_A$. Using
$p_\theta=p_\psi=0$, the equation of motion requires
\begin{equation}\label{string locus 1}
  n_A^{\ 2}=\frac{Q_A}{\sum_B Q_B}\ .
\end{equation}
Similarly, $\delta\vartheta$ equation yields
\begin{equation}\label{string locus 2}
  (\cos^2\vartheta,\sin^2\vartheta)=
  \left(\frac{J_1}{J_1+J_2}\ ,\ \frac{J_2}{J_1+J_2}\right)\ ,
\end{equation}
while $\delta\rho$ equation reduces to
\begin{equation}\label{string locus 3}
  \tanh^2\rho=\frac{J_1+J_2}{\mathcal{H}}\ \ \left(\ \rightarrow\ \
  \cosh^2\rho=\frac{\mathcal{H}}{\mathcal{H}-J_1-J_2} \right)\ .
\end{equation}
Inserting these values back into the expression of the Hamiltonian
(second line of (\ref{tension action})), one obtains
\begin{equation}
  \mathcal{H}=\sum_A Q_A +J_1+J_2\ ,
\end{equation}
which is the BPS energy relation that matches with the
$\frac{1}{16}$-BPS giant gravitons. Now we turn to the equations of
motion for the momentum variables. Those coming from $\delta
p_\rho$, $\delta p_\vartheta$, $\delta p_\theta$ and $\delta p_\psi$
turn out to be trivially satisfied. The equations of motion for
$\delta\vec{Q}$ and $\delta\vec{J}$ reduce to
\begin{equation}
  \dot{\vec{\phi}}=1+\lambda\vec{\phi}^\prime\ \ ,\ \ \
  \dot{\vec{\varphi}}=1+\lambda\vec{\varphi}^\prime\ .
\end{equation}
As in \cite{mmt2}, we choose the gauge $\lambda=0$. This results in
a light-like time evolution for this string, which is
\begin{equation}\label{angle solution}
  \phi^A=t+r^A\sigma\ \ ,\ \ \
  \varphi^a=t+s^a\sigma\ \ .
\end{equation}
As for the $\sigma$ dependent terms, we chose the string to
uniformly wrap the angle directions, as in \cite{mmt2}. Finally, one
has to impose the constraint coming from $\delta\lambda$. With the
above solution (\ref{angle solution}) for the angles, the constraint
is
\begin{equation}\label{string constraint}
  \vec{r}\cdot\vec{Q}+\vec{s}\cdot\vec{J}=0\ .
\end{equation}
This completes the construction of $\frac{1}{16}$-BPS tensionless
string solutions.

\end{appendix}


\begin{thebibliography}{12345}

\bibitem{mst} J. McGreevy, L. Susskind and N. Toumbas, ``Invasion
of the giant gravitons from anti-de Sitter space,'' JHEP {\bf 0006},
008 (2000) [arXiv:hep-th/0003075].

\bibitem{gmt} M.T. Grisaru, R.C. Myers and O. Tafjord, ``SUSY and
goliath,'' JHEP {\bf 0008}, 040 (2000) [arXiv:hep-th/0008015].

\bibitem{hhi} A. Hashimoto, S. Hirano and N. Itzhaki, ``Large
branes in AdS and their field theory dual,'' JHEP {\bf 0008}, 051
(2000) [arXiv:hep-th/0008016].

\bibitem{bbns} V. Balasubramanian, M. Berkooz, A. Naqvi and M.J.
Strassler, ``Giant gravitons in conformal field theory,'' JHEP {\bf
0204}, 034 (2002) [arXiv:hep-th/0107119].

\bibitem{cjr} S. Corley, A. Jevicki and S. Ramgoolam, ``Exact
correlators of giant gravitons from dual N=4 SYM theory,'' Adv.
Theor. Math. Phys. {\bf 5}, 809 (2002) [arXiv:hep-th/0111222].




\bibitem{llm} H. Lin, O. Lunin and J.M. Maldacena, ``Bubbling AdS
space and 1/2 BPS geometries,'' JHEP {\bf 0410}, 025 (2004)
[arXiv:hep-th/0409174].

\bibitem{ber-1} D. Berenstein, ``A toy model for the AdS/CFT
correspondence,'' JHEP {\bf 0407}, 018 (2004)
[arXiv:hep-th/0403110].





\bibitem{surya} N.V. Suryanarayana, ``Half-BPS giants, free
fermions and microstates of superstars,'' JHEP {\bf 0601}, 082
(2006) [arXiv:hep-th/0411145].

\bibitem{sh-to} M.M. Sheikh-Jabbari and M. Torabian,
``Classification of all 1/2 BPS solutions of the tiny graviton
matrix theory,''  JHEP {\bf 0504}, 001 (2005)
[arXiv:hep-th/0501001].



\bibitem{mandal} G. Mandal, ``Fermions from half-BPS
supergravity,'' JHEP {\bf 0508}, 052 (2005) [arXiv:hep-th/0502104].

\bibitem{gmmpr} L. Grant, L. Maoz, J. Marsano, K. Papadodimas and V.S.
Rychkov, ``Minisuperspace quantization of `Bubbling AdS' and free
fermion droplets,'' JHEP {\bf 0508}, 025 (2005)
[arXiv:hep-th/0505079].

\bibitem{dms} A. Dhar, G. Mandal and M. Smedback, ``From
gravitons to giants,''  JHEP {\bf 0603}, 031 (2006)
[arXiv:hep-th/0512312].





\bibitem{mikh} A. Mikhailov, ``Giant gravitons from holomorphic
surfaces,'' JHEP {\bf 0011}, 027 (2000) [arXiv:hep-th/0010206].

\bibitem{kl} S. Kim and K. Lee, ``BPS electromagnetic waves on giant
gravitons,'' JHEP {\bf 0510}, 111 (2005) [arXiv:hep-th/0502007].

\bibitem{ber-2} D. Berenstein, ``Large N BPS states and emergent
quantum gravity,'' JHEP {\bf 0601}, 125 (2006)
[arXiv:hep-th/0507203].

\bibitem{bglm}  I. Biswas, D. Gaiotto, S. Lahiri and S. Minwalla,
``Supersymmetric states of N=4 Yang-Mills from giant gravitons,''
arXiv:hep-th/0606087.

\bibitem{ma-sur} G. Mandal and N.V. Suryanarayana, ``Counting
1/8-BPS dual giants,'' arXiv:hep-th/0606088.

\bibitem{ast} M. Ali-Akbari, M. M. Sheikh-Jabbari and M. Torabian,
``Tiny graviton matrix theory/SYM correspondence: analysis of BPS
states,'' arXiv:hep-th/0606117.



\bibitem{my-ta} R.C. Myers and O. Tafjord, ``Superstars and giant
gravitons,'' JHEP {\bf 0111}, 009 (2001) [arXiv:hep-th/0109127].

\bibitem{bcs} K. Behrndt, A.H. Chamseddine and W.A. Sabra, ``BPS black
holes in N=2 five dimensional AdS supergravity,'' Phys. Lett. B{\bf
442}, 97 (1998) [arXiv:hep-th/9807187].




\bibitem{gu-re-1} J.B. Gutowski and H.S. Reall,
``Supersymmetric AdS$_5$ black holes,'' JHEP {\bf 0402}, 006 (2004)
[arXiv:hep-th/0401042].

\bibitem{gu-re-2} J.B. Gutowski and H.S. Reall,
``General supersymmetric AdS$_5$ black holes,'' JHEP {\bf 0404}, 048
(2004) [arXiv:hep-th/0401129].

\bibitem{cclp1} Z.-W. Chong, M. Cvetic, H. Lu and C.N. Pope,
``Five-dimensional gauged supergravity black holes with independent
rotation parameters,'' Phys. Rev. D{\bf 72}, 041901 (2005)
[arXiv:hep-th/0505112].

\bibitem{cclp2} Z.-W. Chong, M. Cvetic, H. Lu and C.N. Pope,
``General non-extremal rotating black holes in minimal
five-dimensional gauged supergravity,'' Phys. Rev. Lett. {\bf 95},
161301 (2005) [arXiv:hep-th/0506029].

\bibitem{klr} H.K. Kunduri, J. Lucietti and H.S. Reall,
``Supersymmetric multi-charge AdS$_5$ black holes,'' JHEP {\bf
0604}, 036 (2006) [arXiv:hep-th/0601156].




\bibitem{kmmr} J. Kinney, J.M. Maldacena, S. Minwalla and S. Raju,
``An index for 4 dimensional super conformal theories,''
arXiv:hep-th/0510251.

\bibitem{brs} M. Berkooz, D. Reichmann and J. Simon, ``A Fermi
surface model for large supersymmetric AdS(5) black holes,''
arXiv:hep-th/0604023.

\bibitem{silva} P.J. Silva, ``Thermodynamics at the BPS bound for
black holes in AdS,'' arXiv:hep-th/0607056.





\bibitem{cglp}  M. Cvetic, G.W. Gibbons, H. Lu and C.N. Pope,
``Rotating black holes in gauged supergravities; thermodynamics,
supersymmetric limits, topological solitons and time machines,''
arXiv:hep-th/0504080.

\bibitem{kp} A. Kostelecky and M. Perry,
``Solitonic black holes in gauged N=2 supergravity,'' Phys. Lett.
B{\bf 371}, 191 (1996) [arXiv:hep-th/9512222].




\bibitem{mmt1}  D. Mateos, T. Mateos and P.K. Townsend,
``Supersymmetry of tensionless rotating strings in AdS$_5 \times$
S$^5$, and nearly-BPS operators,''  JHEP {\bf 0312}, 017 (2003)
[arXiv:hep-th/0309114].

\bibitem{mmt2} D. Mateos, T. Mateos and P.K. Townsend, ``More on
supersymmetric tensionless rotating strings in AdS$_5 \times$
S$^5$,'' arXiv:hep-th/0401058.




\bibitem{gu-sa} J.B. Gutowski and W. Sabra, ``General supersymmetric
solutions of five-dimensional supergravity,'' JHEP {\bf 0510}, 039
(2005) [arXiv:hep-th/0505185].

\bibitem{cvetic} M. Cvetic, M.J. Duff, P. Hoxha, J.T. Liu, H. Lu,
J.X. Lu, R. Martinez-Acosta, C.N. Pope, H. Sati and T.A. Tran,
``Embedding AdS black holes in ten-dimensions and
eleven-dimensions,'' Nucl. Phys. B{\bf 558}, 96 (1999)
[arXiv:hep-th/9903214].



\bibitem{ma-to} D. Mateos and P.K. Townsend, ``Supertubes,''
Phys. Rev. Lett. {\bf 87}, 011602 (2001) [arXiv:hep-th/0103030].

\bibitem{pa-mar} B.C. Palmer and D. Marolf, ``Counting
supertubes,'' JHEP {\bf 0406}, 028 (2004) [arXiv:hep-th/0403025].

\bibitem{bho} D. Bak, Y. Hyakutake and N. Ohta, ``Phase moduli
space of supertubes,'' Nucl. Phys. B{\bf 696}, 251 (2004)
[arXiv:hep-th/0404104].



\bibitem{bhko} D. Bak, Y. Hyakutake, S. Kim and N. Ohta, ``A
geometric look on the microstates of supertubes,'' Nucl. Phys. B{\bf
712}, 115 (2005) [arXiv:hep-th/0407253].



\bibitem{dghw} A. Dabholkar, J.P. Gauntlett, J.A. Harvey and
D. Waldram, ``Strings as solitons and black holes as strings,''
Nucl. Phys. B{\bf 474}, 85 (1996) [arXiv:hep-th/9511053].

\bibitem{lu-ma} O. Lunin and S.D. Mathur, ``Metric of the multiply
wound rotating string,'' Nucl. Phys. B{\bf 610}, 49 (2001)
[arXiv:hep-th/0105136].

\bibitem{ch-oh}  J. Cho and P. Oh, ``Super D-helix,''
Phys. Rev. D{\bf 64}, 106010 (2001) [arXiv:hep-th/0105095].

\bibitem{mnt} D. Mateos, S. Ng and P.K. Townsend, ``Supercurves,''
Phys. Lett. B{\bf 538}, 366 (2002) [arXiv:hep-th/0204062].





\bibitem{aw-to} M. Awada and P. K. Townsend, ``N=4 Maxwell-Einstein
supergravity in five-dimensions and its SU(2) gauging,'' Nucl. Phys.
B{\bf 255}, 617 (1985).

\bibitem{ma-nu} J.M. Maldacena and C. Nunez, ``Supergravity
description of field theories on curved manifolds and a no go
theorem,'' Int. J. Mod. Phys. A{\bf 16}, 822 (2001)
[arXiv:hep-th/0007018].






\end{thebibliography}
\end{document}